\documentclass[structabstract,a4paper]{aa}  
\usepackage{txfonts}
\usepackage{xspace}
\usepackage{natbib}
\usepackage{graphicx}
\usepackage{url}
\begin{document}
\title{Analyzing X-ray pulsar profiles:\\ geometry and beam pattern of A\,0535+26}
   \author{I.~Caballero \inst{1}
          \and  U.~Kraus \inst{2}
          \and A.~Santangelo \inst{3} 
          \and M.~Sasaki \inst{3} 
          \and P.~Kretschmar \inst{4}
        }
        \institute{CEA Saclay, DSM/IRFU/SAp --UMR AIM (7158) 
          CNRS/CEA/Universit\'{e} P. Diderot, Orme des Merisiers 
          Bat. 709, 91191 Gif-sur-Yvette, France 
          \email{isabel.caballero@cea.fr}
          \and Institut f\"ur Physik und Technik, Universit\"at Hildesheim, 
          Marienburger Platz 22, 31141, Hildesheim, Germany
          \and Institute f\"ur Astronomie und Astrophysik, Kepler Center 
          for Astro and Particle Physics, Eberhard Karls Universit\"at 
          T\"ubingen, Sand 1, 72076 T\"ubingen, Germany 
          \and ISOC, European Space Astronomy Centre (ESAC), PO Box 78, 
          28691 Villanueva de la Ca\~nada, Spain}             
        
        \date{}
\titlerunning{Geometry and beam pattern of A\,0535+26}

        \abstract
        {}
        {We applied a decomposition method to the energy dependent 
            pulse profiles of the accreting binary pulsar A\,0535+26, in order to 
            identify the contribution of the two magnetic poles of the neutron 
            star and to obtain constraints on the geometry of the system and 
            on the beam pattern. }
        {We analyzed pulse profiles obtained from RXTE observations in the 
            X-ray regime. Basic assumptions of the method are 
            that the asymmetry observed in the pulse profiles is caused by 
            non-antipodal magnetic poles and that the emission regions have 
            axisymmetric beam patterns.   
        }
        {Constraints on the geometry of the pulsar
          and a possible solution of the beam pattern are given. 
          We interpreted the reconstructed beam pattern in terms of a 
          geometrical model of a hollow column plus a halo of scattered 
          radiation on the neutron star surface, which includes relativistic 
          light deflection.}
        {}
        
        \keywords{X-rays: binaries -- Pulsars: individual: A\,0535+26
        }
        \maketitle
   \section{Introduction}
   The Be/X-ray binary A\,0535+26 was discovered by \textsl{Ariel V} in 1975 
   \citep{rosenberg75} during a giant outburst. The system consists of a 
neutron star orbiting the optical companion HDE\,245770 on an eccentric orbit 
($e=0.47\pm0.02$) of orbital period $P_{\mathrm{orb}}=111.1\pm 0.3$\,d 
\citep{finger06}. The source presents quiescent X-ray emission with a 
luminosity of $L_{\mathrm{X}}\lesssim10^{35-36}\,\mathrm{erg}\,\mathrm{s}^{-1}$, 
sometimes interrupted by ``normal'' outbursts 
($L_{\mathrm{X}}\approx10^{37}\,\mathrm{erg}\,\mathrm{s}^{-1}$) linked to the 
periastron passages of the neutron star, and less frequent ``giant'' outbursts 
($L_{\mathrm{X}}>10^{37}\,\mathrm{erg}\,\mathrm{s}^{-1}$) of longer duration and 
less clearly related to the orbital phase 
\citep[see, e.g.,][]{giovanelli92,kend94,finger96}. 
The system presents two cyclotron resonant scattering 
features at $E\sim45$\,keV and $E\sim100$\,keV, from which a magnetic 
field strength of $B\sim 4\times10^{12}$\,G 
is inferred (\citealt{kend94,grove95,kretschmar05,caballero07}). 

Pulsations are observed to have a period of $P_{\mathrm{spin}}\sim$103\,s, 
typically with spin-up during stronger outbursts and spin-down during 
quiescent periods\footnote{see, e.g., the results of \textsl{Fermi}-GBM 
monitoring at\newline 
\url{http://gammaray.nsstc.nasa.gov/gbm/science/pulsars/}}.
The pulse profile evolves from a complex profile at lower energies
to a simpler, two-peaked structure at higher energies. This
behavior is observed in several accreting X-ray pulsars.
Similar to other sources \citep[e.g.,][]{staubert80}, 
individual pulses show strong 
pulse-to-pulse variations, while the average pulse profile 
is rather stable, with slower variations over the course of
an outburst \citep{caballero08a}.

The basic concept of pulsed emission is well understood. Pulsed emission originates 
in regions close to the magnetic poles of a rotating neutron star with the 
magnetic axis misaligned with respect to the rotation axis.  
In contrast, physical modeling of the pulsed emission turns out to be a complex task. 
Many processes are in fact involved in modeling pulse profiles,
from the modeling of the emission regions and their local emission pattern to the 
formation of the pulse profiles seen by a distant observer. Comparison of model 
calculations with observations has been performed for instance by 
\citet{wang81}, \citet{meszaros85}, and \cite{leahy91}.

A proper model calculation should include relativistic light deflection, 
which has a significant effect on the pulse shape\footnote{The 
importance of relativistic  light deflection in model calculations can be 
visualized in 
\url{http://www.spacetimetravel.org/xpulsar06/xpulsar06.html}} 
\citep{riffert88}. For slowly rotating neutron stars, the metric around a 
neutron star can be approximated by the Schwarzschild metric 
(see, e.g., \citealt{pechenick83}). Due to the strong gravitational field 
around the neutron star, the X-rays will be observed at red-shifted energies.  
Geometrical models of filled and hollow accretion columns of accreting 
neutron stars, including relativistic light deflection, were 
computed in \cite{kraus01} and \cite{kraus03}. 
These models give the beam pattern or energy-dependent flux of one 
emission region as a function of the angle, as seen by a distant observer. 
Introducing the rotation of the pulsar and its geometry, i.e., the orientation 
of the rotation axis with respect to the direction of observation and the 
location of the two poles, the pulsed emission from each of the two poles 
(single-pole pulse profiles) observed by a distant observer can be modeled. 
The sum of the single-pole contributions gives the total pulse profile.

An alternative method of analyzing pulse profiles is to start from the 
observed pulse profiles and, based on symmetry considerations,  
decompose the pulse profile into single-pole contributions. 
This is then transformed into the visible section of the beam pattern.
This method has been successfully applied to the accreting X-ray binary 
pulsars Cen\,X-3\, Her\,X-1 and EXO\,2030+375 (\citealt{kraus96}, \citealt{blum00}, 
\citealt{sasaki10} respectively) and is applied in this work to A\,0535+26. 
Preliminary results were presented in \cite{caballero08c}.

\section{Method}
\subsection{Assumptions}
A major goal of this work is to reconstruct the beam pattern of the neutron star in the 
binary system A\,0535+26. By beam pattern we refer to the emission of one magnetic pole, as seen 
by a distant observer, as a function of the angle $\theta$ between the direction of observation 
and the magnetic axis (see Fig.~\ref{fig:kraus95_1}). 
The basic assumption made in this work, which is often adopted in model calculations, is 
that the emission regions at the magnetic poles are axisymmetric. 
The magnetic axis is therefore a symmetry axis of the emission from the neutron star. 
Under this assumption, the beam pattern from one pole is only a function of angle $\theta$, and therefore 
the pulsed emission of each magnetic pole or single-pole pulse profile 
has to be symmetric. One of the symmetry points 
will be the instant when the magnetic axis is closest to the line of sight, 
and the other symmetry point will be half a period later, with the magnetic 
axis pointing away from the observer. 

In the case of an ideal dipole field, the sum of the two single-pole pulse profiles will give a 
symmetric pulse profile because the two single-pole pulse profiles have the same symmetry points. 
However, the observed pulse profiles of accreting X-ray pulsars are generally asymmetric.  
To explain the asymmetry observed in the pulse profiles, we assume 
a dipole magnetic field with a small offset from an ideal dipole field 
(see Fig.~\ref{fig:pulsar_geo}).
A small deviation of one of the magnetic poles from the antipodal position 
makes the symmetry points of the two single-pulse profiles different, 
causing the asymmetry in the total pulse profile.

Another assumption made in this work is that the two emission regions are the same, 
i.e., have the same beam pattern. This implies that each of the two poles
will make one section of the same beam pattern visible. Depending 
on the geometry of the neutron star and the angle of observation, 
those two sections will have coincident parts in some cases.
This assumption was tested with the accreting pulsars Cen X-3 
\citep{kraus96} and Her X-1 \citep{blum00}. In these cases an 
overlapping region was found, in agreement with the assumption
of two equal emission regions.  
\subsection{Decomposition into single-pole pulse profiles}
The first step of the analysis is to express the 
original pulse profile as a Fourier series. The total pulse profile $F$ 
is written as

\begin{equation}
F(\Phi)=\frac{1}{2}u_{0}+\displaystyle\sum_{k=1}^{n/2-1} [u_{k}\cos(k\Phi)
+v_{k}\sin(k\Phi)]+u_{n/2}\cos(\frac{n}{2}\Phi)
\label{eq:fourier}
\end{equation}
where $n$ is the number of bins of the original pulse profile, $\Phi$ 
the phase, and $u_{k}$, $v_{k}$ are given by
\begin{eqnarray}\label{eq:u}
u_{k}=\frac{1}{\pi}\int_{-\pi}^{+\pi}F(\Phi)\cos(k\Phi)d\Phi\\
v_{k}=\frac{1}{\pi}\int_{-\pi}^{+\pi}F(\Phi)\sin(k\Phi)d\Phi
\label{eq:v}
\end{eqnarray}
Equation~\ref{eq:fourier} gives a valid representation of the original pulse
profile at all phases if the Fourier transform of $F$ approaches zero 
as the frequency approaches $n/2$.

The single-pole pulse profiles $f_{1}$ and $f_{2}$ are described
by the following symmetric functions:

\begin{equation}
f_{1}(\Phi)=\frac{1}{2}c_{0}+\displaystyle\sum_{k=1}^{n/2} c_{k}\cos[k(\Phi-\Phi_{1})]
\label{eq:f1}
\end{equation}

\begin{equation}
f_{2}(\Phi)=\frac{1}{2}d_{0}+\displaystyle\sum_{k=1}^{n/2} d_{k}\cos\{k[\Phi-(\Phi_{2}+\pi)]\},
\label{eq:f2}
\end{equation}
where $\Phi_{1}$ and $\Phi_{2}$ are the symmetry points of $f_{1}(\Phi)$ and 
$f_{2}(\Phi)$, respectively. Formally, a decomposition of
$F$ into two symmetric functions exists for every choice of their
symmetry points $\Phi_{1}$ and $\Phi_{2}$. For convenience,
we use the parameter $\Delta := \pi -(\Phi_{1}-\Phi_{2})$,
which represents the azimuthal displacement of one pole with respect 
to the antipodal position (see Sec.~\ref{sec:profiles_to_bp}). 
All formal decompositions will be contained
in the parameter space $\Phi_{1}-\Delta$, with $0\le\Phi_{1}\le\pi$
and $0\le\Delta\le\pi/2$. Once the formal decompositions are calculated, 
the following criteria are applied to reduce the number of decompositions 
to physically meaningful ones: 

\begin{itemize}
\item \textsl{positive criterion}: both $f_{1}(\Phi)$ and $f_{2}(\Phi)$ 
must be positive, since they represent photon fluxes;
\item \textsl{non-ripple criterion}: the single-pole contributions 
$f_{1}(\Phi)$ and $f_{2}(\Phi)$ should not be much more complicated than 
the original pulse profile. Individual pulse profiles with many peaks 
that cancel out in the sum are not accepted;
\item the same symmetry points $\Phi_{1}$ and $\Phi_{2}$ must give 
valid decompositions in all energy bands. 
\end{itemize}
Once a possible decomposition is found, the symmetry points
for each of the two poles $\Phi_{1}$ and $\Phi_{2}$, and the 
parameter $\Delta$, related to the position of the emission regions 
on the neutron star, are determined.   
\subsubsection{From single-pole pulse profiles to beam pattern}
\label{sec:profiles_to_bp}

\begin{figure}
\centering
\includegraphics[height=8cm,angle=180,bb=138 720 542 408,clip=]{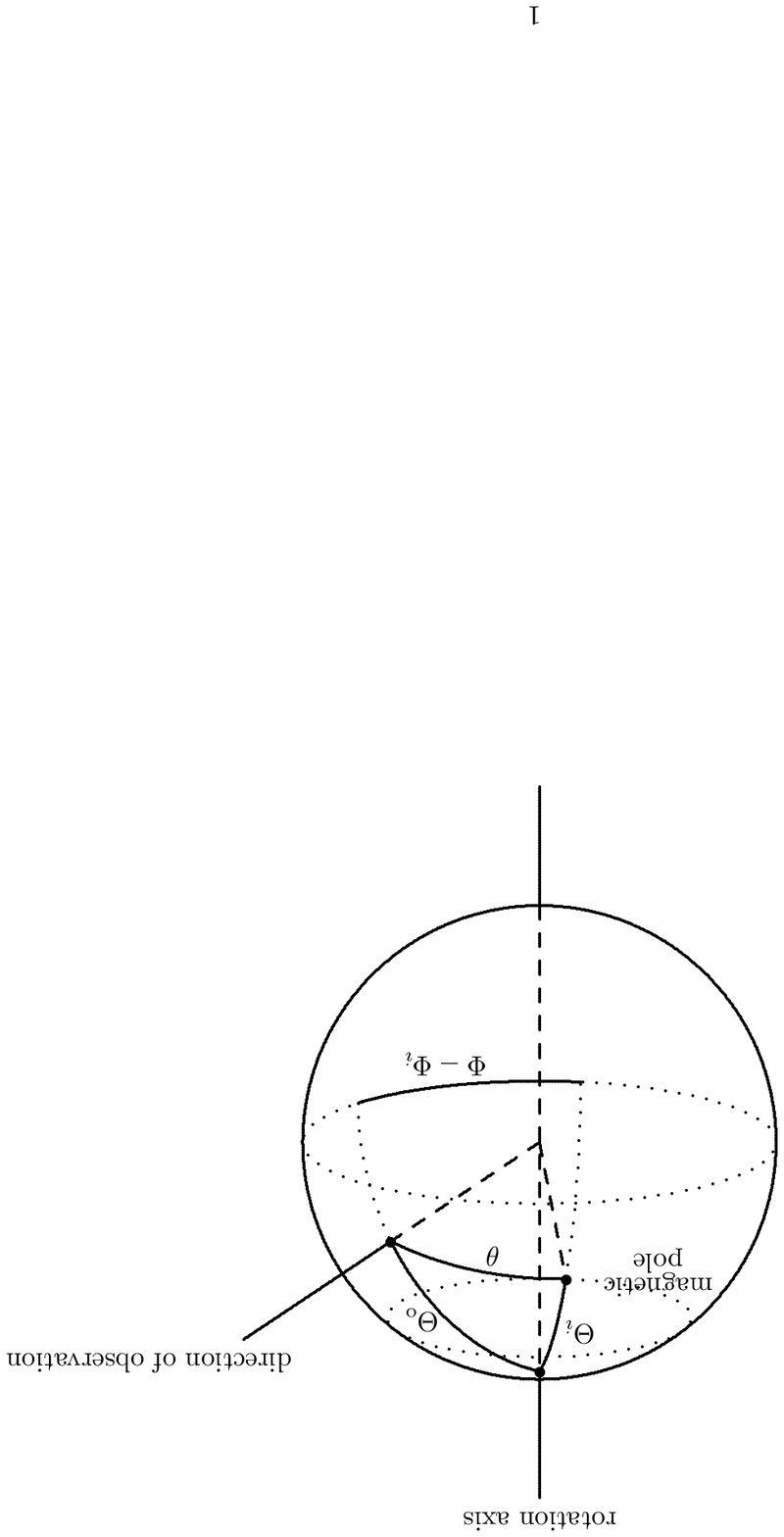}
\caption{Schematic view of a rotating neutron star from \cite{kraus95}. A spherical coordinate 
system is used, with the rotation axis as polar axis. $\Theta_{i}$ is the 
polar angle of the \textsl{i}th pole. $\Theta_{0}$
is the polar angle of the direction of observation. The angle $\theta$ 
between the magnetic pole and the direction of observation changes with 
the rotation angle $\Phi$.}
\label{fig:kraus95_1}
\end{figure}

In Fig.~\ref{fig:kraus95_1} a schematic view of a rotating neutron 
star is shown. A spherical coordinate system is used with the rotation 
axis as polar axis.  
As explained above, the beam pattern is assumed to be axisymmetric, 
hence to only depend on the angle between the direction of 
observation and the magnetic axis $\theta$. The value of $\theta$ changes 
with the rotation angle $\Phi$. 
Depending on the position of the poles with 
respect to the rotation axis and depending on the direction of observation
with respect to the magnetic axis, we only observe a section 
of the beam pattern for each pole. 

Applying the cosine formula to the spherical triangle in 
Fig.~\ref{fig:kraus95_1}, we obtain $\theta$ as a function 
of the phase $\Phi$:

     \begin{equation}
     \cos\theta=\cos\Theta_{0}\cos\Theta_{i}+\sin\Theta_{0}\sin\Theta_{i}\cos(\Phi-\Phi_{i})
     \label{eq:theta}
     \end{equation}  
where $\Theta_{0}$ is the polar angle of the direction of observation, 
$\Theta_{i}$ the polar angle of the \textsl{i}th pole, and
$\Phi_{i}$ one symmetry point for the \textsl{i}th pole.

\begin{figure}
\centering
\includegraphics[height=8cm,angle=180,bb=138 720 542 408,clip=]{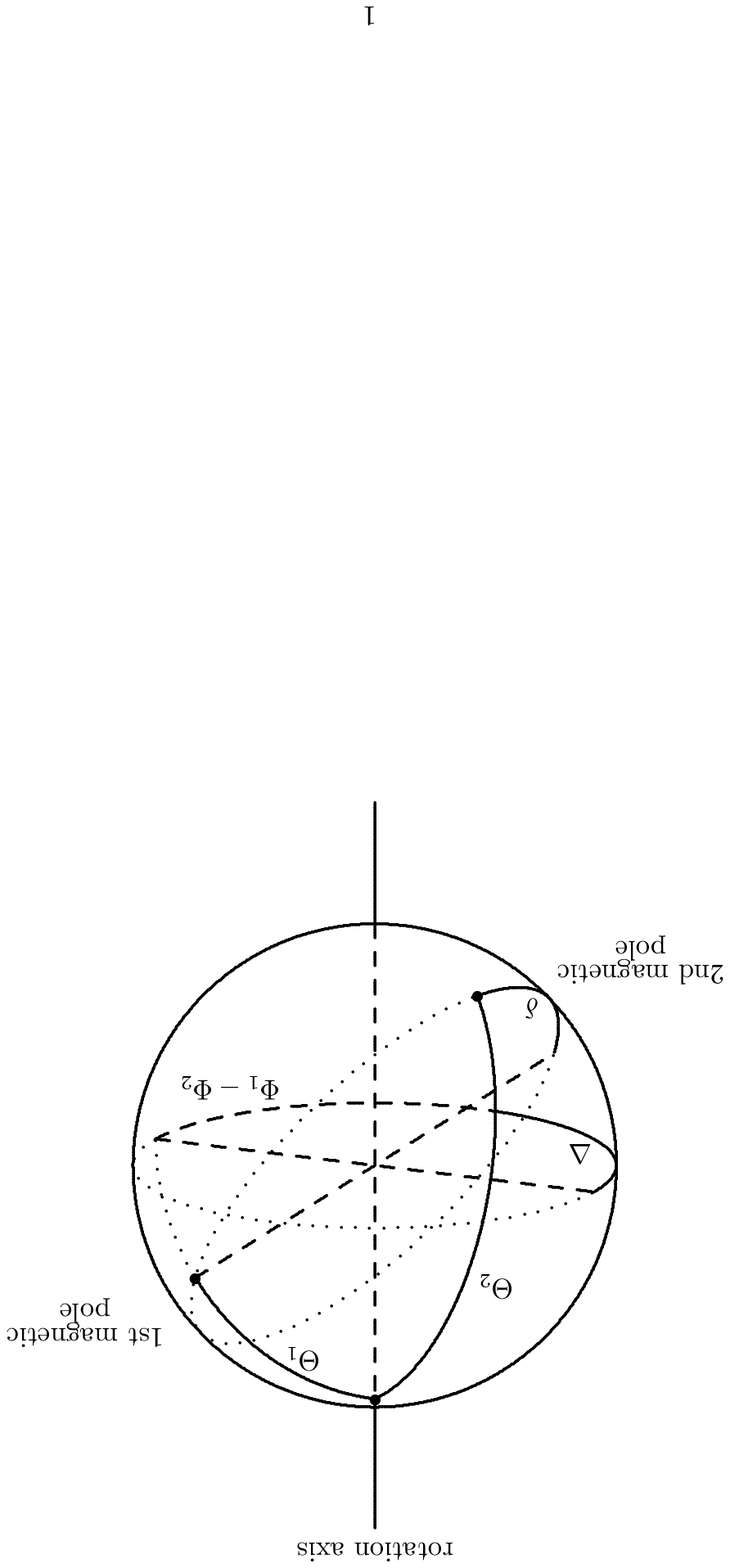}
\caption{Intrinsic geometry of the pulsar. With the rotation
axis as polar axis, the magnetic poles are located at polar 
angles $\Theta_{1}$ and $\Theta_{2}$. The angular distance $\delta$
between the second magnetic pole and the point that is antipodal 
to the first magnetic pole gives the deviation from an ideal dipole
field. Figure from \cite{kraus95}.}
\label{fig:pulsar_geo}
\end{figure}

The intrinsic pulsar geometry is shown in Fig.~\ref{fig:pulsar_geo}.
A complete description of the pulsar can be given in terms
of the polar angles $\Theta_{1}$ and $\Theta_{2}$, and the difference
in their azimuthal angles $\Phi_{1}-\Phi_{2}=\pi-\Delta$. 
The angular distance $\delta$ between the location 
of the second magnetic pole and the point that is antipodal to the 
first magnetic pole can be used as a measure for the deviation 
from an ideal dipole field. From Fig.~\ref{fig:pulsar_geo} comes

\begin{equation}
\cos\delta=-\cos\Theta_{2}\cos\Theta_{1}+\sin\Theta_{2}\sin\Theta_{1}\cos\Delta.
\label{eq:delta}
\end{equation}

Considering the beam pattern as a function of  $\cos\theta$ and 
the single-pole pulse profiles as functions of $\cos(\Phi-\Phi_{i})$,   
there is no distortion between the two functions because the relation 
between $\cos\theta$ and $\cos(\Phi-\Phi_{i})$ is linear (Eq.~\ref{eq:theta}). 
Therefore, once we have the single-pole pulse profiles, by plotting them as a function 
of $\cos(\Phi-\Phi_{i})$ we obtain two different sections of the same beam pattern. 
These two sections are different because we see a different interval 
of $\theta$ for each of the two magnetic poles. Depending on the geometry, these two 
ranges of $\theta$ can overlap, so there will be an overlapping region 
in the beam pattern. In this region, at an instant $\Phi$, the first pole will be seen 
at an angle $\theta$. At another instant $\tilde{\Phi}$, the second pole 
will be seen at the same angle $\theta$. We can use Eq.~\ref{eq:theta} 
to express the relation between $\Phi$ and $\tilde{\Phi}$ in terms of 
the geometric parameters
\begin{equation}
\cos(\Phi-\Phi_{1})=\frac{\cot\Theta_{0}(\cos\Theta_{2}-\cos\Theta_{1})}{\sin\Theta_{1}}+\frac{\sin\Theta_{2}}{\sin\Theta_{1}}\cos(\tilde{\Phi
}-\Phi_{2})
\label{eq:Phi}
\end{equation}
that we write as

\begin{equation}
\cos(\Phi-\Phi_{1})=a+b\cos(\tilde{\Phi}-\Phi_{2}),~~~b>0
\label{eq:ab}
\end{equation}
where the parameter $a$ represents the shift between the two single-pole 
pulse profiles.  

The sections that the single-pole pulse profiles have in common can be found 
by representing them as a function of $\cos(\Phi-\Phi_{i})$, so the values of 
$a$ and $b$ can be determined. Once this is done, both sections of the beam 
pattern can be represented as a function of the same variable $q$, defined as

\begin{equation}
q:=\frac{\cos\theta-\cos\Theta_{0}\cos\Theta_{1}}{\sin\Theta_{0}\sin\Theta_{1}}. 
\end{equation}

Using Eqs.~\ref{eq:theta} and \ref{eq:ab}:

\begin{equation}
\cos(\Phi-\Phi_{1})=q,
\label{eq:q}
\end{equation}

\begin{equation}
\cos(\tilde{\Phi}-\Phi_{2})=(q-a)/b. 
\label{eq:qab}
\end{equation}

Since the relation between $q$ and $\cos\theta$ is linear, by plotting 
the two single-pole pulse profiles as a function of $q$ we obtain the 
total visible part of the beam pattern without distortion. 
If the values of $a$ and $b$ have been determined, by comparing  
Eqs.~\ref{eq:Phi} and \ref{eq:ab} we obtain two equations relating the 
three geometric parameters $\Theta_{0}$, $\Theta_{1}$, and $\Theta_{2}$. 
They can be solved for $\Theta_{1}$ and $\Theta_{2}$ as a function 
of $\Theta_{0}$:

\begin{equation}
\tan\Theta_{1}=\frac{-2a\,\tan\Theta_{0}}{(a\,\tan\Theta_{0})^2+b^2-1},
\label{eq:theta1}
\end{equation}

\begin{equation}
\tan\Theta_{2}=\frac{b\,\tan\Theta_{1}}{a\,\tan\Theta_{0}\,\tan\Theta_{1}+1}.
\label{eq:theta2}
\end{equation}

To obtain the location of the magnetic poles $\Theta_{1}$ and 
$\Theta_{2}$, an independent determination of the direction of observation $\Theta_{0}$ is necessary. 
If $\Theta_{0}$ is known, the position of the two magnetic poles can be obtained from 
Eqs.~\ref{eq:theta1} and \ref{eq:theta2}. The pulsar geometry is then completely 
determined: the location of the poles $\Theta_{1}$ and $\Theta_{2}$, their displacement from the antipodal   
position $\Delta$, and the direction of observation $\Theta_{0}$. 
Once the geometric parameters of the system are known, the single-pole pulse profiles can 
be expressed as a function of $\theta$ using Eq.~\ref{eq:theta}. The observable 
part of the beam pattern is in this way completely reconstructed. 
Further details can be found in \cite{kraus95}.

\subsection{Application to A 0535+26}
\label{sec:application_A 0535+26}
\begin{figure}
\centering
\includegraphics[height=8cm,angle=0]{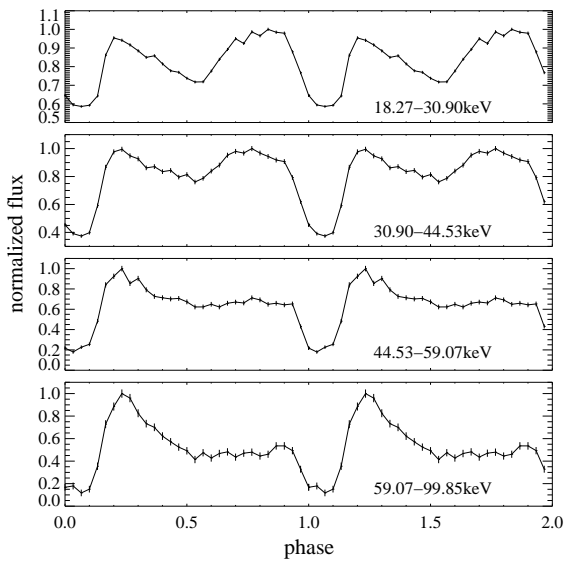}
\caption{Example of A\,0535+26 original pulse profiles used in the decomposition 
analysis, obtained from RXTE-HEXTE observations performed during its 
August/September 2005 normal outburst (MJD 53614.66), in the energy ranges 
(18.27--30.90)\,keV, (30.90--44.53)\,keV, (44.53--59.07)\,keV, 
(59.07--99.85)\,keV from top to bottom.  The fluxes have been normalized to 
unity. Two pulse cycles are shown for clarity.}
\label{fig:original}
\end{figure}

Our analysis is based on RXTE-HEXTE observations of A\,0535+26 
during its August/September 2005 normal outburst. 
Pointed observations performed between 
MJD 53614.66--53615.6 were used (observation IDs 91085-01-01-03, 
-01-01-05, -01-02-00, -01-02-01, -01-02-02, -01-02-03, 
for a total exposure time of $\sim$20\,ks). 
We extracted RXTE-HEXTE background-substracted light curves in the energy 
ranges  18.27--30.90\,keV, 30.90--44.53\,keV, 44.53--59.07\,keV, 
59.07--99.85\,keV. 
After applying barycentric and orbital correction to the light 
curves (ephemeris from \citealt{finger06}), we folded them with 30 bins, using the best pulse 
period from \citet{caballero08a}. An example of the energy-dependent pulse profiles is given in Fig.~\ref{fig:original}. 
The pulse profiles selected for the analysis correspond to the main part of the outburst, as those profiles 
remain stable during the outburst and also during historical observations (see, e.g., \citealt{frontera85,kend94,finger96}). 
The flux during the observations was  
$F_{(3-50)\,\mathrm{keV}}\approx(1.66-1.85)\times10^{-8}\,\mathrm{erg}\,\mathrm{cm}^{-2}\,\mathrm{s}^{-1}$, 
giving a luminosity of 
$L_{(3-50)\,\mathrm{keV}}\approx(0.79-0.885)\times10^{37}\,\mathrm{erg}\,\mathrm{s}^{-1}$
assuming a distance of $d=2\,$kpc \citep{steele98}. The flux was determined from 
PCA and HEXTEspectra, that we modeled with a power law with a high-energy cutoff plus two 
Gaussian absorption lines at $\sim45\,$ and $\sim100\,$keV. Details 
of the spectral and timing analysis of the observations can be found in \citet{caballero07,caballero08a}. 

\subsubsection{Search for acceptable decompositions}\label{sec:single_pole_profiles_A 0535+26}
\begin{figure*}
\centering
\includegraphics[width=0.49\textwidth]{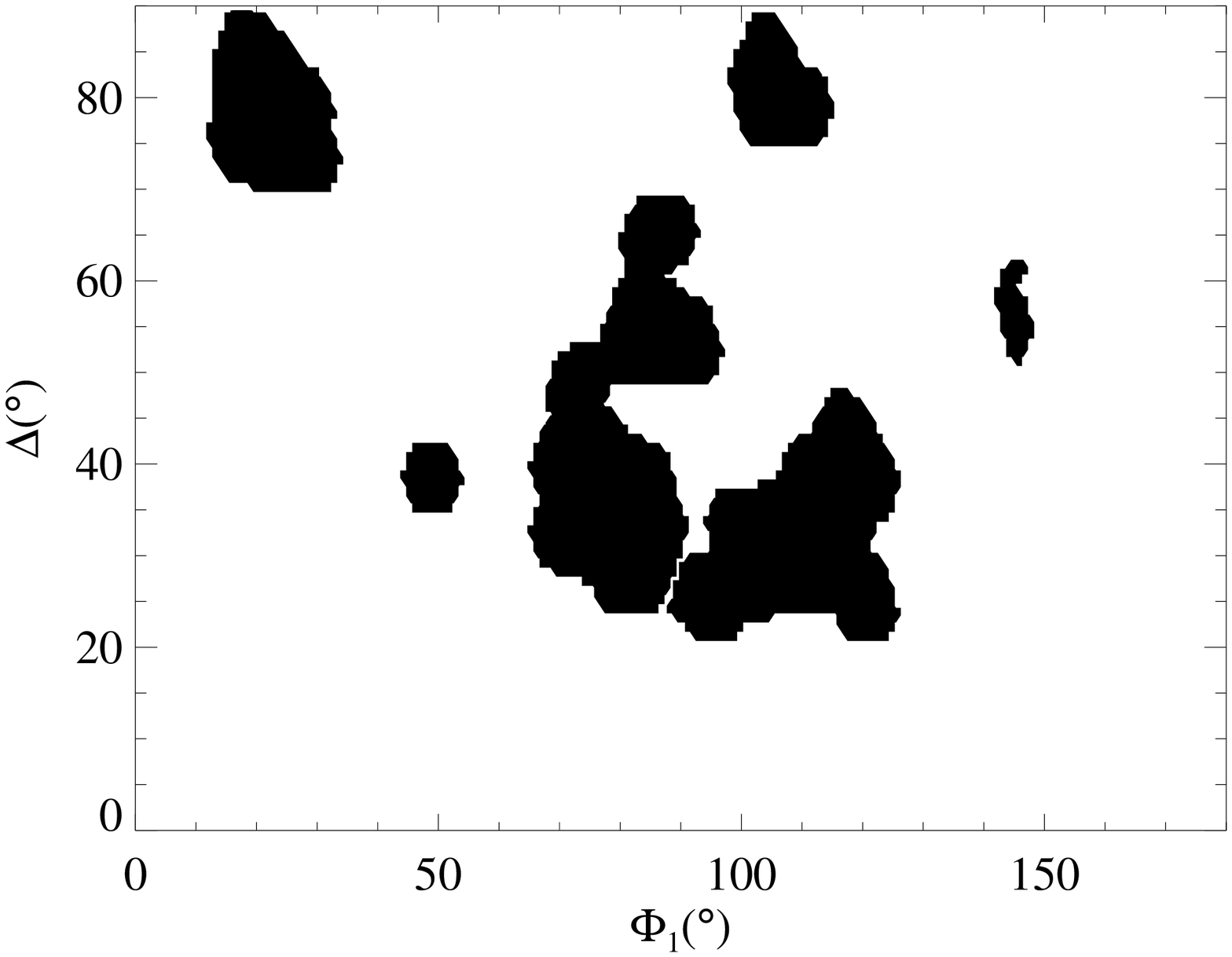}
\includegraphics[width=0.49\textwidth]{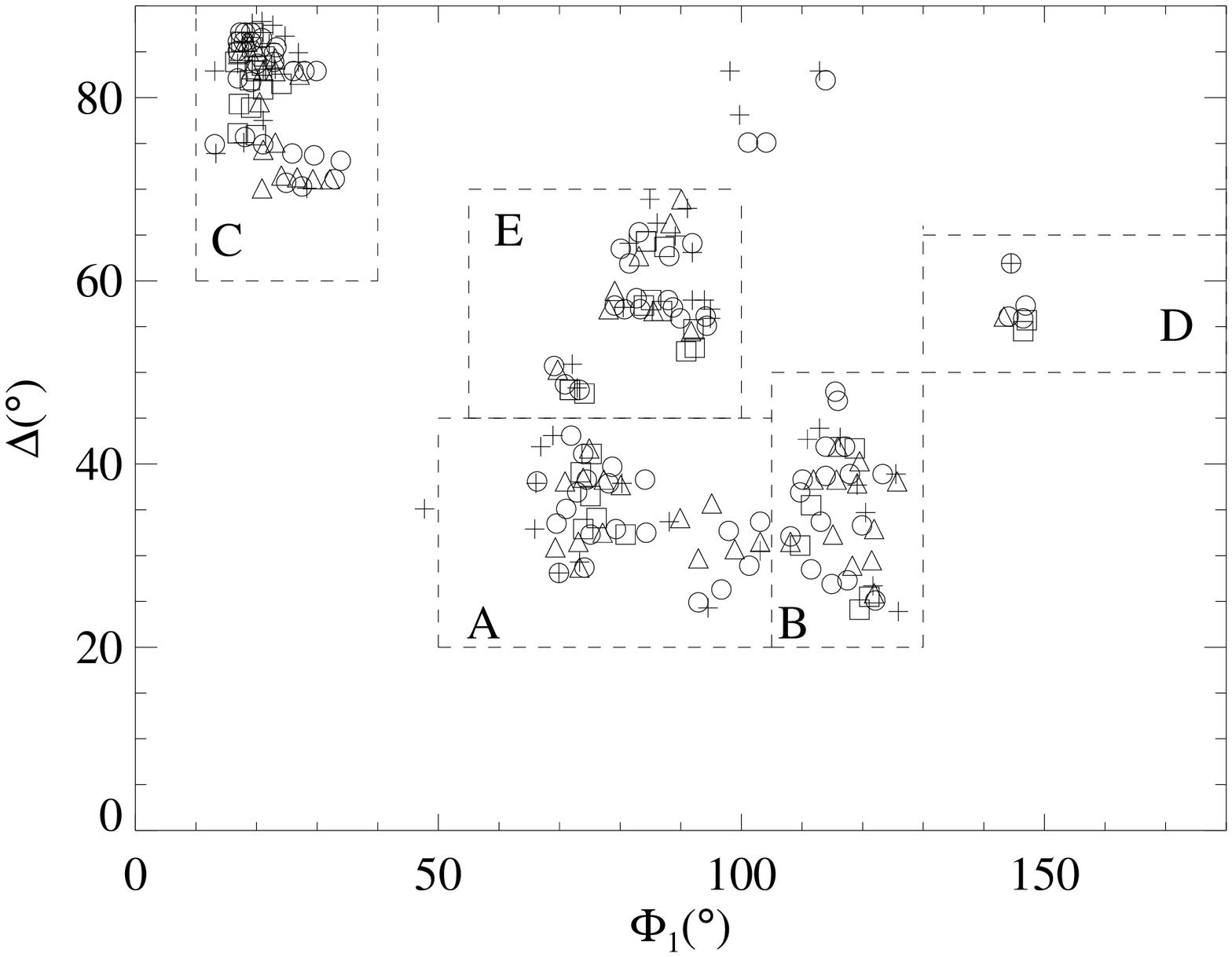}
\caption{\textsl{Left}: Result of applying the \textsl{non-negative} criterion: 
decompositions outside the black area are discarded. \textsl{Right}: 
Highest ranked decompositions according to the \textsl{non-ripples} 
criterion grouped in five regions 
$A$, $B$, $C$, $D$, and $E$. 
The different symbols represent different energy ranges: 
+ (18.3--30.9)\,keV,
$\times$ (30.9--44.5)\,keV , 
$\triangle$ (44.5--59.1)\,keV, 
$\Box$  (59.1--99.8)\,keV. }
\label{fig:contour_decompositions}
\end{figure*}

The energy-dependent pulse profiles are expressed as Fourier series using an FFT algorithm.  
Twenty Fourier coefficients out of the 30 sampled points are enough to 
properly describe the original profiles. The highest frequency terms are not 
considered in order to avoid aliasing (see e.g. \citealt{press92}). 
These functions are then written as the sum of two symmetric functions  
$f_{1}(\Phi)$ and $f_{2}(\Phi)$. The condition $F(\Phi)$=$f_{1}(\Phi)$+$f_{2}(\Phi)$  
allows us to obtain the coefficients $c_{k}$ and $d_{k}$ (Eqs.~\ref{eq:f1},~\ref{eq:f2}) 
as a function of $u_{k}$ and $v_{k}$ (Eqs.~\ref{eq:u}, ~\ref{eq:v}). 
To search for physically meaningful decompositions, the $\Phi_{1}-\Delta$ 
parameter space is divided in $1\,^{\circ}\times1\,^{\circ}$ boxes. All the
formal decompositions are represented in this plane. 
First we apply the \textsl{non-negative criterion}. This criterion is 
more restrictive when the minimum of the pulse profile is close to zero. 
Knowledge of the background level is particularly important in this 
step. A small amount of negative flux is allowed to account 
for the $\sim1$\,\% uncertainty of the HEXTE background    
(\citealt{rothschild98}). The \textsl{non-negative} criterion reduces 
the number of acceptable decompositions considerably, which are mainly driven by the 
higher energy pulse profiles  (E=59.1--99.8\,keV) that have a pulsed 
fraction of $\sim$79\,\%. The result of applying   
this criterion, which requires each decomposition to be valid at all energy 
ranges, can be seen in the $\Phi_{1}-\Delta$ plane in 
Fig.~\ref{fig:contour_decompositions} (left).  The black regions indicate 
where positive decompositions have been found. We limit the search of 
physically acceptable decompositions to those regions.

The \textsl{non-ripples criterion} is then applied. 
The original pulse profiles in the considered energy ranges 
have two main peaks. We expect the single-pole contributions not 
to have many more peaks than the original one. 
The number of peaks of the single-pole pulse profiles are counted. 
A quality function is defined as the inverse of the total number of peaks. With this 
method we obtain the decompositions sorted according to their quality
function. Higher values of the quality function correspond to decompositions
that are simpler and not much more complicated than the original pulse profile.

For each decomposition, there is a certain region in the 
$\Phi_{1}$-$\Delta$ plane that contains other qualitatively similar 
decompositions. To handle the large amount of possible 
decompositions that have to be studied, similar ones are grouped 
together into types. A decomposition belongs to a certain type if its square 
deviation with respect to the chosen representative decomposition of that type 
is smaller  than a certain $\chi^{2}=10^{-3}$.  
Grouping them into types considerably reduces the number of decompositions 
that are left to study, since we only consider one 
representative of each type. For instance, in the $18.27-30.90\,$keV range, 1863 
decompositions are grouped into 286 types. Figure \ref{fig:contour_decompositions} (right) 
shows where the highest ranked profile representatives were found, grouped into types. 
We examined all these decompositions in different energy ranges,
dividing the parameter space in five regions $A$, $B$, $C$, $D$, and $E$, 
shown in Fig. ~\ref{fig:contour_decompositions} (right). 

\begin{figure}
\includegraphics[width=0.5\textwidth]{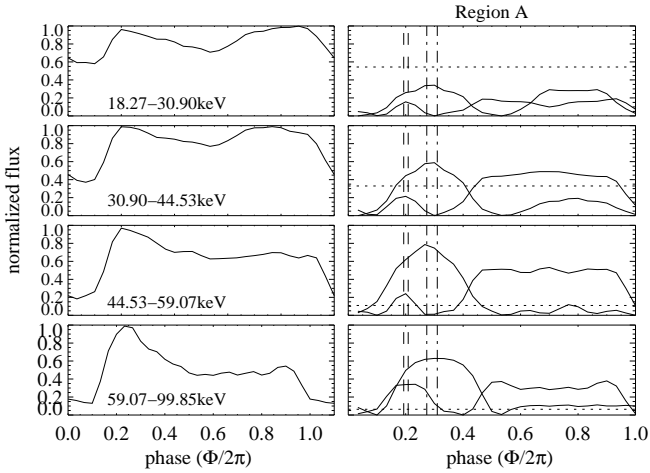} 
\caption{\textsl{Left}: original pulse profiles described with a Fourier series, 
normalized to unity in four different energy ranges. \textsl{Right}: best decomposition of the 
original pulse profiles in two symmetric functions, corresponding to 
region $A$. The vertical lines indicate the ranges for the symmetry points 
for both single-pole pulse profiles, $\Phi_{1}$ (dashed lines) and 
$\Phi_{2}$ (dash-dotted lines). The dotted horizontal lines represent the 
unmodulated flux left available to distribute between the two functions. }
\label{fig:decomp}
\end{figure}

\begin{figure}
\includegraphics[width=0.5\textwidth]{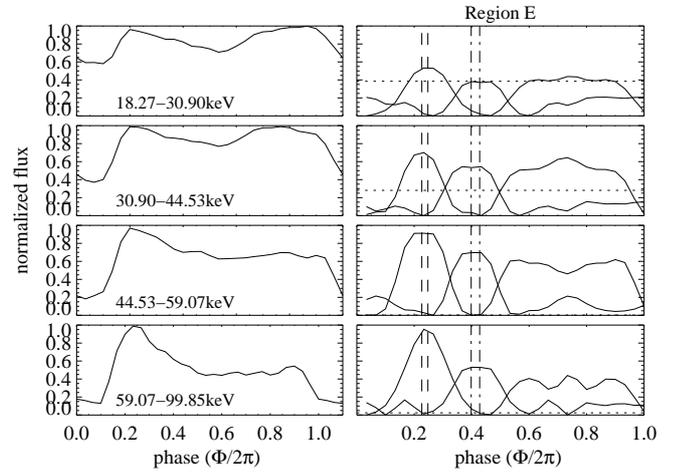} 
\caption{\textsl{Left}: original pulse profiles described with a Fourier series, 
normalized to unity in four different energy ranges. \textsl{Right}: ``second best'' decomposition of 
the original pulse profiles in two symmetric functions, corresponding to 
region $E$. Dashed and dash-dotted lines have the same meaning as in 
Fig.~\ref{fig:decomp}.}
\label{fig:decompE}
\end{figure}

As discussed in detail below, by combining different energy ranges and different 
observations, we find the best decomposition of the original pulse profiles 
for the parameters of Region $A$. They are shown in Fig.~\ref{fig:decomp}, with 
the remaining unmodulated flux available to distribute 
between the two symmetric functions that cannot be determined from the 
decomposition. The minima of the two symmetric functions have been shifted
to zero, so that the sum of the two symmetric functions plus
the unmodulated flux reproduces the original pulse profile. 
As expected, the single-pole pulse profiles at all 
energy ranges are not much more complicated than the sum. The symmetry points 
obtained for this best decomposition, calculated as the average from the 
different energy ranges, are given in Table~\ref{tab:phidelta}.
We take 5\,$^{\circ}$ as the estimate for the uncertainty in the symmetry points,  
which is the approximate range in region $A$ for which the 
decompositions are similar. 

\begin{table}[!h]
  \caption[Symmetry points $\Phi_{1}$, $\Phi_{2}$ and $\Delta$ for best 
        decomposition]
        {Symmetry points and azimuthal displacement of one pole with 
        respect to the antipodal position for the best decomposition of 
        the A\,0535+26 pulse profiles.}
  \label{tab:phidelta} \centering
\begin{tabular}{|c | c | c|}\hline
Pole 1 & Pole 2 &  $\Delta=\pi-(\Phi_{1}-\Phi_{2})$\\\hline
$\Phi_{1}=72\degr\pm5\degr$ & $\Phi_{2}=285\degr\pm5\degr$ & $\Delta=33\degr\pm5
\degr$\\
$\Phi_{1}+\pi=252\degr\pm5\degr$& $\Phi_{2}+\pi=105\degr\pm5\degr$  & \\\hline
  \end{tabular}
\end{table}

The decompositions in Region $C$ are discarded because they present a very 
strong anti-correlation in the main peaks which seems artificial, not 
expected from two independent emission regions. The single-pole
pulse profiles are generated independently, so they should not 
have features that match exactly and cancel out in the sum. Decompositions in Regions 
$B$ and $D$ have also been discarded, because the single-pole pulse 
profiles present an anti-correlation in many small features that cancel out 
in the sum, which is also not expected from two independent emission regions. In 
Region $E$ we find the ``second best" decomposition for 
$\Phi_{1}=82\,\degr\pm5\,\degr$ and $\Delta=63\,\degr\pm2\,\degr$. The 
single-pole pulse profiles are not much more complicated than the sum, and 
present an energy evolution similar to that of the total pulse profile. 
However, we favor the solution in Region $A$ compared to Region $E$ because 
the single-pole pulse profiles in Region $E$ present a strong anti-correlation 
in the main features, not expected from a physical point of view.
Another argument for rejecting decompositions in Regions $C$, $D$, and $E$ is that
they all present higher values of $\Delta$. Under the assumption of 
slightly displaced magnetic poles, lower values of $\Delta$ are more likely
to be real. This was the case in the analysis of the accreting pulsars 
Cen X-3 \citep{kraus96} and Her X-1 \citep{blum00}, where the best 
decompositions were found for low values of $\Delta$. A further argument 
against decompositions in Regions $B$, $C$, and $D$ emerges in the 
reconstruction of the beam pattern from the single-pole contributions 
(see Sect.~\ref{sec:geo_beam_pattern_A 0535+26}). This argument does not emerge 
for Region $E$, and therefore the beam pattern obtained for Region $E$ is  
also discussed below. 

\subsubsection{From single-pole pulse profiles to geometry and beam pattern}
\label{sec:geo_beam_pattern_A 0535+26}

In the previous section (\ref{sec:single_pole_profiles_A 0535+26}) we 
obtained a decomposition of the pulse profiles of A\,0535+26 in two single-pole components. 
We therefore have the pulse profile for each magnetic pole for a given energy range 
as a function of the rotation angle $\Phi$. As shown in Sect.~\ref{sec:profiles_to_bp}, using Eq.\ref{eq:theta}, 
by representing the single-pole pulse profiles as a function 
of $\cos(\Phi-\Phi_{i})$, we obtain two sections of the same beam pattern. 
We performed this transformation to the single-pole pulse profiles 
of A\,0535+26. In contrast to the accreting pulsars Her\,X-1 and Cen\,X-3, an overlapping region is not observed 
for A\,0535+26. 

However, under the assumption of two identical emission regions, the two 
sections can almost be connected to each other, with a small gap 
in between in the beam pattern that remains unobservable to us. 
This means that, owing to the geometry of the neutron star and to its rotation, 
we are observing two different sections of the same beam pattern. 
However, unlike the case of an overlapping region, we cannot determine the parameters $a$ and $b$ of 
Eq.~\ref{eq:ab} from a fit. But we can estimate their values. 
We combined the sections of the beam pattern in all energy ranges using 
different values of $a$, which represents the shift between the two single-pole pulse profiles. 
The best estimate for the parameter $a$ is $a=2.2\pm0.1$. 
In the case of antipodal poles, $b=1$ \citep{kraus95}. 
Since we are assuming a small distortion, $b$ should be close to 1, so we 
make the assumption $b=1$. 

\begin{figure}
\centering
\includegraphics[width=0.5\textwidth]{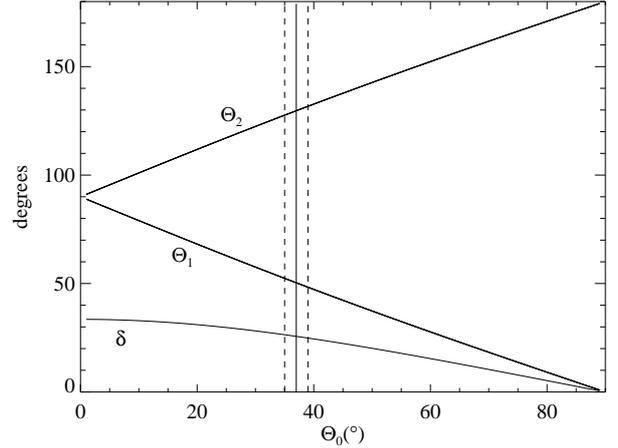}
\caption{Constraints on the pulsar geometry. The vertical lines
        indicate the orbital inclination (solid line) and its error 
        range (dashed lines).}
\label{fig:geo}
\end{figure}

With these estimates for $a$ and $b$ plus the direction of observation
$\Theta_{0}$, it is possible to obtain the location of the poles $\Theta_{1}$ 
and $\Theta_{2}$. The angular distance between the location of the second 
pole and the point that is antipodal to the first pole $\delta$
can be estimated using Eq.~\ref{eq:delta}. In Fig.~\ref{fig:geo} the 
constraints on the geometry of the pulsar are shown. The values of 
$\Theta_{1}$, $\Theta_{2}$ and $\delta$ are represented for all possible 
values of the direction of observation $\Theta_{0}$. The mass transfer from the 
optical companion is expected to align the rotation axes of the binary system on 
a shorter timescale than the system's lifetime. Assuming that the 
rotation axis of the neutron star is perpendicular to the orbital plane,  
the inclination of the system is identical to the direction of 
observation $i=\Theta_{0}$. \cite{giovannelli07} determined an inclination 
for the system of $i=37\,\pm2\,^{\circ}$. We can therefore obtain the location 
of the poles $\Theta_{1}$ and $\Theta_{2}$. Table~\ref{tab:geo} gives the 
estimated values we find for the position of the magnetic poles and the 
offset.

\begin{table}
  \caption[Geometrical parameters of A\,0535+26 from the decomposition]
  {Geometrical parameters of A\,0535+26 from the decomposition 
    analysis. See Fig.~\ref{fig:pulsar_geo} for 
    the definition of the angles.} 
  \label{tab:geo} \centering
  \begin{tabular}{|c | c |c|}\hline
    Pole 1 & Pole 2 & offset from ideal dipole\\\hline
    $\Theta_{1}\approx50\degr$& $\Theta_{2}\approx130\degr$ & $\delta\approx25\degr$
    \\\hline
  \end{tabular}
\end{table}

\begin{figure*}
\centering
\includegraphics[height=7cm,angle=0]{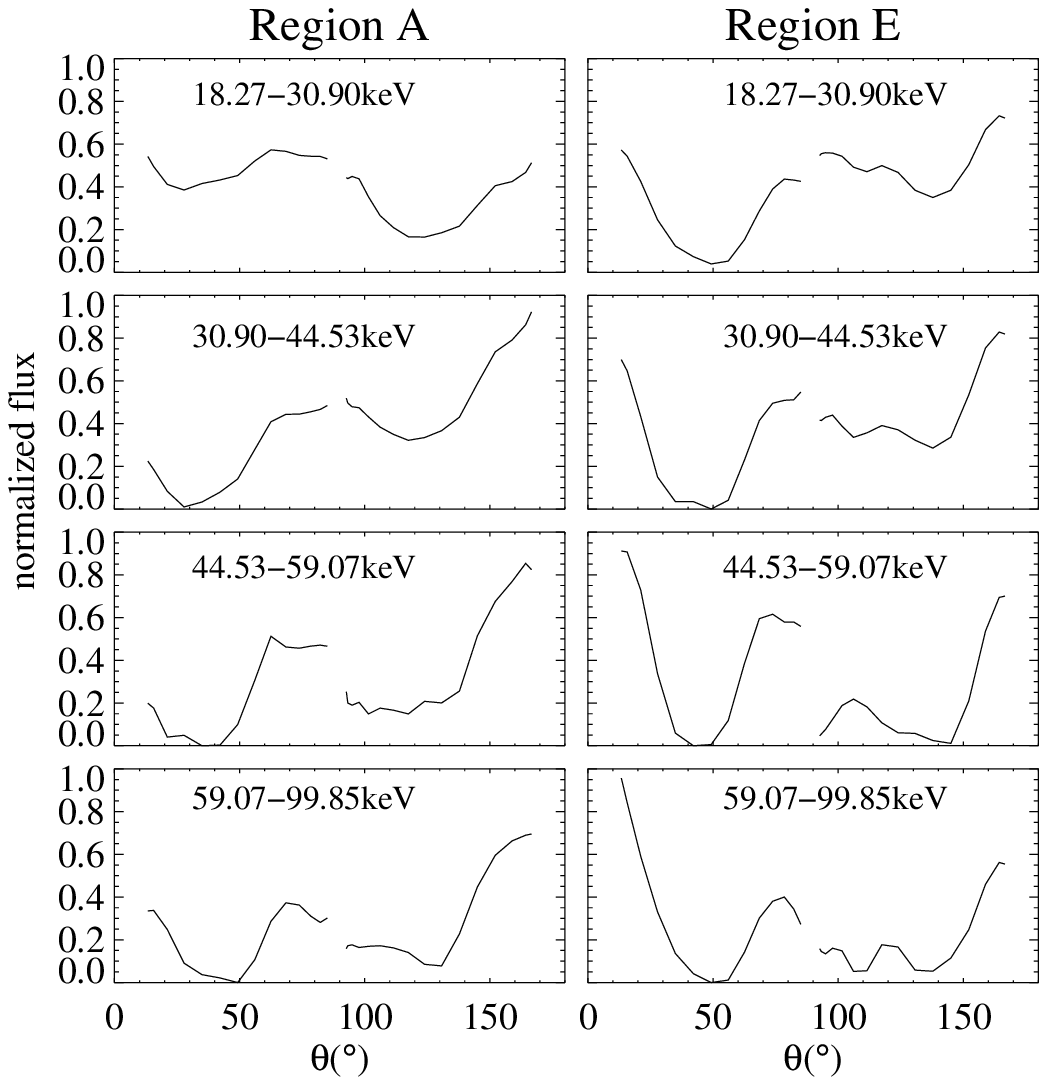}
\includegraphics[height=7cm,angle=0]{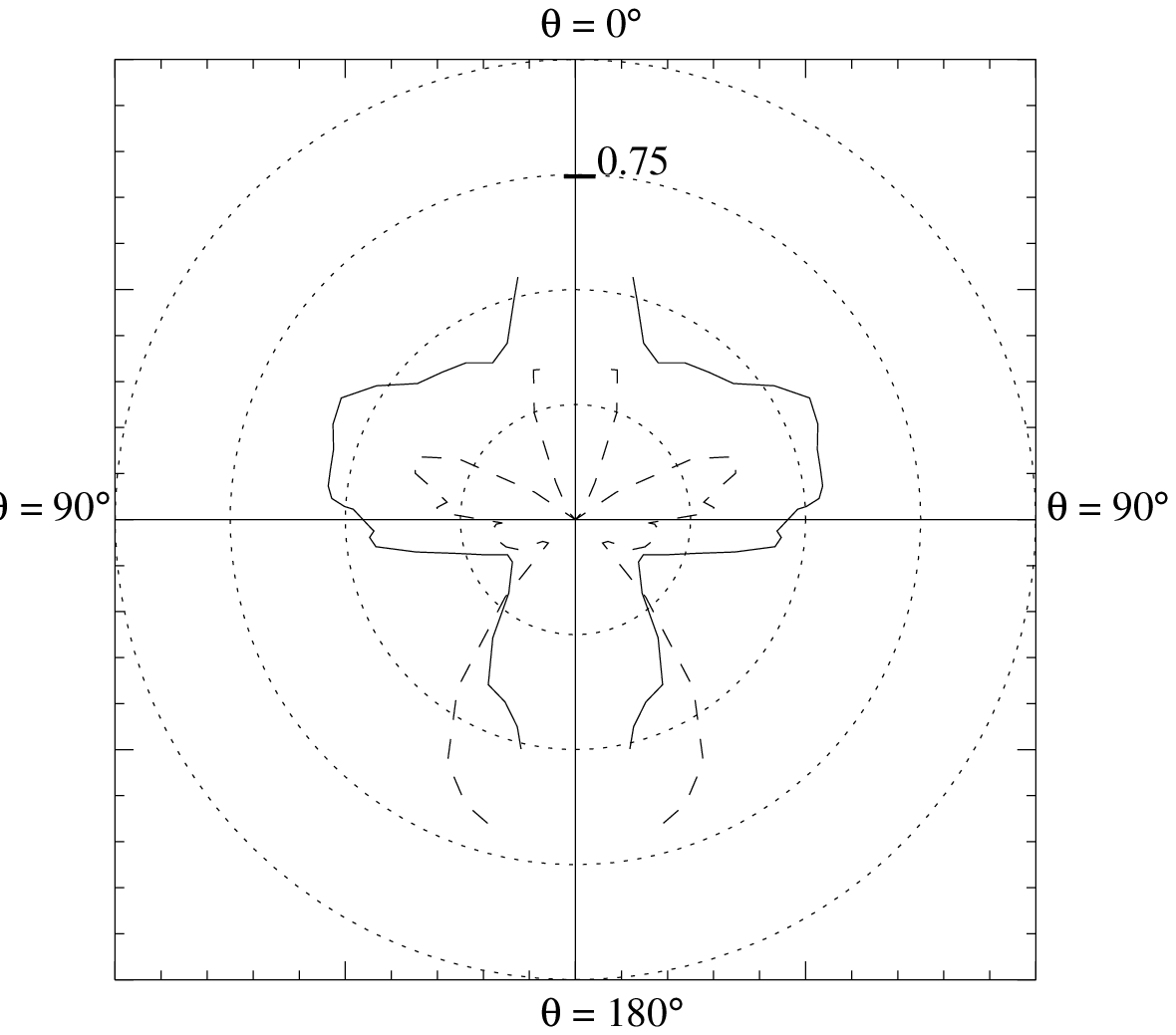}
\caption{ \textsl{Left}: Reconstructed beam pattern of A\,0535+26 or emission of one magnetic
pole as a function $\theta$, as seen by a distant observer. The  
beam pattern is shown in linear representation for Regions $A$ (left panels) and 
$E$ (right panels) for different energy ranges.  
\textsl{Right}: reconstructed beam pattern for Region $A$ in polar representation, for 
the energy ranges 18.27-30.90\,keV (solid line) and 59.07--99.85\,keV (dashed line). 
$\theta=0\degr$ means that the distant observer is looking down onto the magnetic pole, 
and $\theta=180\degr$ means that the observer looks at the pole from the antipodal position. 
}
\label{fig:bp}
\end{figure*}

It is then possible to plot the reconstructed sections of the beam 
pattern as a function of $\theta$. The two sections of the beam 
pattern are reconstructed for $\theta\in(13^{\circ},87^{\circ})$ and 
$\theta\in(93^{\circ},167^{\circ})$. Figure \ref{fig:bp} (left panels)
shows the reconstructed beam pattern or emission of one magnetic pole as a function of $\theta$, 
with $\theta=0\degr$ meaning the distant observer is looking down onto the magnetic pole, 
and $\theta=180\degr$ meaning the observer looks at the pole from the antipodal position.   
Figure \ref{fig:bp} (right) shows the same beam pattern in polar representation. 

For regions $B$, $C$, and $D$, it was not possible to connect the two single pole 
contributions, providing a further argument for discarding them.  
For the ``second best" decomposition (Region $E$), it has been possible to 
connect the reconstructed sections of the beam pattern in a similar way 
as for Region $A$, also suggesting that the two emission regions are the same. 
We can connect them with the same values 
of $a$ and $b$ as in Region $A$. This implies the same values for the 
polar angles of the magnetic poles $\Theta_{1}$ and $\Theta_{2}$. 
The value of $\Delta$ is higher than in region $A$, so that the offset from an ideal 
dipole field is also higher, $\delta\approx48\,\degr$. In Fig.~\ref{fig:bp} left (right panels) 
the beam patterns for the best decompositions in region $E$ are represented for comparison. They differ 
slightly from those in Region $A$, but the main features of the two functions 
are similar. The single-pole pulse profiles also had very similar shapes, 
just different symmetry points. 

\subsubsection{Interpretation and modeling of the beam pattern}
\label{sec:interpretation}
A characteristic feature of the reconstructed beam pattern is
a minimum observed in the flux between $\theta\approx30^{\circ}-40^{\circ}$ 
(Fig.~\ref{fig:bp} left). 
This feature is present at all energies, leading us to believe that
it could be related to the geometry of the accretion. Filled column models give a beam pattern in which the flux decreases 
at low values of $\theta$, corresponding to the instant when the observer 
looks along the accretion stream \citep{kraus03}. Introducing a 
hollow column plus a halo created on the neutron star surface 
around the column walls from scattered radiation emitted from the walls 
would explain the increase in flux at low values of $\theta$ and the minimum 
as $\theta$ increases when the observer looks directly into the column. 

Another characteristic feature of the reconstructed beam patterns at all energies 
is a steep increase in flux at high values of $\theta$ ($\theta>120^{\circ}$, meaning the 
observer looks at one pole of the neutron star from the antipodal position). 
This can be due to gravitational light bending, making the emission from one pole 
visible from all directions and brightest from the antipodal position. Such a steep 
increase in flux at high values of $\theta$ has been observed in model calculations, 
producing a maximum in the flux at 
$\theta=180^{\circ}$ \citep{kraus03}. This feature has also been observed 
in the reconstructed beam patterns of other X-ray pulsars (Her\,X-1 and EXO\,2030+375, see  
\citealt{blum00}, \citealt{sasaki10}). 

To obtain estimates on the size of the accretion column, we applied a phenomenological 
model of a hollow column to the case of A\,0535+26. The model has been computed for values of $\theta\in[0,40]\degr$, 
to try to reproduce the minimum observed in the reconstructed beam patterns of A\,0535+26 for low values of $\theta$ 
(Fig.~\ref{fig:bp}, left). The applied model describes medium-luminosity pulsars for which
the local luminosity per pole is on the order of

\begin{equation}
  L_\ast =
  \frac{\alpha}{4\sqrt{2}} L_{\rm Ed} \frac{\sigma_{\rm T}}{\sigma_{\rm s}}=
  4.5\cdot 10^{36} {\rm \frac{erg}{s}} 
       \left(\frac{\alpha}{0.1\,{\rm rad}}\right )
       \left( \frac{M}{1.4\,M_\odot} \right )
       \frac{\sigma_{\rm T}}{\sigma_{\rm s}}.
\label{eq:lum}
\end{equation}
Here, $\alpha$ is the half opening angle of the base of the accretion
funnel, $L_{\rm Ed}= 4\pi c m_{\rm p} G M / \sigma_{\rm T}$ the
Eddington luminosity, $M$ the mass of the neutron star, $\sigma_{\rm T}$ 
the Thomson scattering cross section, and $\sigma_{\rm s}$ the
magnetic scattering cross section averaged over direction and
frequency. In this case a radiative shock is expected to occur at a
height that is a small fraction of the stellar radius. The shock
separates a region of freely falling plasma above from a settling
region of nearly stagnant plasma below, with radiation mainly
originating in the shock and the settling region and escaping from the
sides of the subshock column or mound \citep{basko_sunyaev_76}. 

The modeling is performed as in \cite{kraus03} and includes a
small accretion column, the formation of a halo on
the neutron star surface, an accretion funnel delimited by magnetic
field lines, magnetic scattering in the accretion stream, and
relativistic light deflection. This is an extension of the purely
geometric hollow column models presented in \cite{kraus01}. A detailed
study of this model will be presented in \cite{kraus10}. 

Beam patterns were computed in the range $\theta < 40^\circ$
for accretion onto a neutron star
with the canonical values of mass $M=1.4\, M_\odot$ and radius $r_{\rm n} =
10$~km. According to the observations of A\,0535+26, we take an
asymptotic luminosity per pole of 
$L_\infty = 8\times 10^{36}\,\mathrm{erg}\,\mathrm{s}^{-1}$,
corresponding to a local luminosity per pole of 
$L = L_\infty / (1-r_{\rm s}/r_{\rm n}) = 1.4\times 10^{37}\,\mathrm{erg}\,\mathrm{s}^{-1}$ 
that is close to the limiting luminosity in Eq.~\ref{eq:lum} for 
typical values of $\alpha$ as studied below.
The radiation is emitted from the side of the column below a radiative
shock set at $r_{\mathrm{t}}=10.5$~km, and the local emission is taken to be an
isotropic blackbody at the effective temperature of the column
wall. This simple emission pattern is based on more detailed studies
that have shown that the bulk of the radiation is expected to escape
from the side of the accretion funnel (Wang and Frank 1981) and that,
in static columns, the local emission in many circumstances is close
to isotropic (M\'{e}sz\'{a}ros and Nagel 1985).

\begin{figure}
\centering
\includegraphics[width=0.45\textwidth]{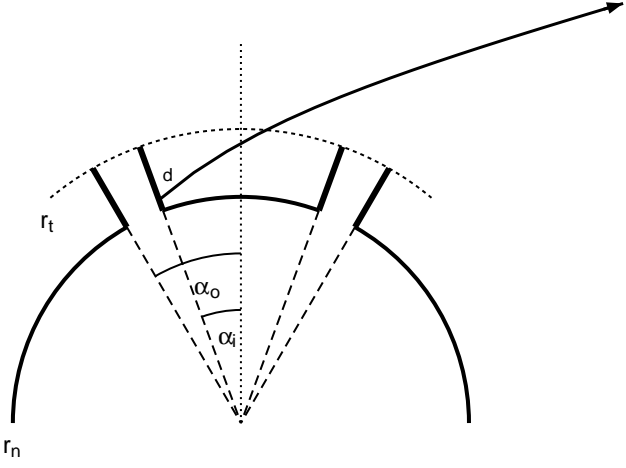}
\caption{Geometrical model of hollow column \citep{kraus01}.}
\label{fig:geo_model}
\end{figure}

The inner and outer half opening angles $\alpha_\mathrm{i}$ and 
$\alpha_\mathrm{o}$ of the hollow accretion funnel (see 
Fig.~\ref{fig:geo_model}) were varied as listed in Table~\ref{tab:models}. 
The table also lists the effective temperature of the column wall and the 
density that the accreting material has at the base of the free fall column. 
These parameters follow from the assumed geometry and the prescribed 
luminosity as described in \cite{kraus03}.
The effective temperature of the column walls, when assuming an 
isotropic black body, is derived from $\sigma T_{\mathrm{eff}}^{4}A_{\mathrm{co}}=L$, 
where $\sigma$ is the Stefan–-Boltzmann constant, $A_{\mathrm{co}}$ the emitting area, 
and $L$ the local luminosity per pole. The free-fall density in the plasma 
rest frame at the base of the free fall column (at $r=r_{\mathrm{n}}$) is taken to be homogeneous 
over the polar cap, and 
derives from $\rho_{0}c^{3}\beta\gamma(\gamma-1)A_{\mathrm{cap}}=L$, where 
$\beta=v/c$ and $\gamma=1/\sqrt{1-\beta^{2}}$ are evaluated at $r=r_{\mathrm{n}}$
(see \citealt{kraus10} for more details). 

\begin{figure}
\centering
 \includegraphics[width=0.5\textwidth]{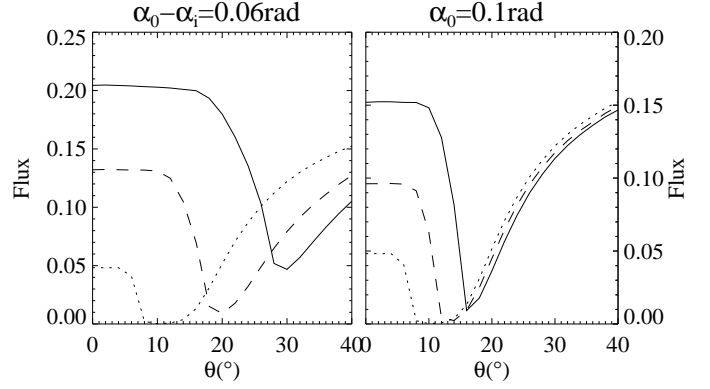}
\caption{Beam pattern models for a hollow column with a halo of scattered 
  radiation on the neutron star surface. $\theta$ has the same meaning as in Fig.~\ref{fig:bp}. 
The minimum is produced when observer looks in the direction of the accretion stream.  \textsl{Left}: 
fixed column thickness 
  $\alpha_{\mathrm{o}}-\alpha_{\mathrm{i}}=0.06$\,rad and different opening 
  angles $\alpha_{\mathrm{o}}=0.2$\,rad (solid line), 
  $\alpha_{\mathrm{o}}=0.15$\,rad (dashed line),  
  $\alpha_{\mathrm{o}}=0.1$\,rad (dotted line).  
  \textsl{Right}: fixed outer opening angle $\alpha_{\mathrm{i}}=0.1\,$rad and 
  different column thickness
  $\alpha_{\mathrm{o}}-\alpha_{\mathrm{i}}=0.02\,$rad (solid line), 
  $\alpha_{\mathrm{o}}-\alpha_{\mathrm{i}}=0.04\,$rad (dashed line) and 
  $\alpha_{\mathrm{o}}-\alpha_{\mathrm{i}}=0.06\,$rad (dotted line). }
\label{fig:diff_models}
\end{figure}

\begin{figure}
  \centering
  \includegraphics[width=0.5\textwidth]{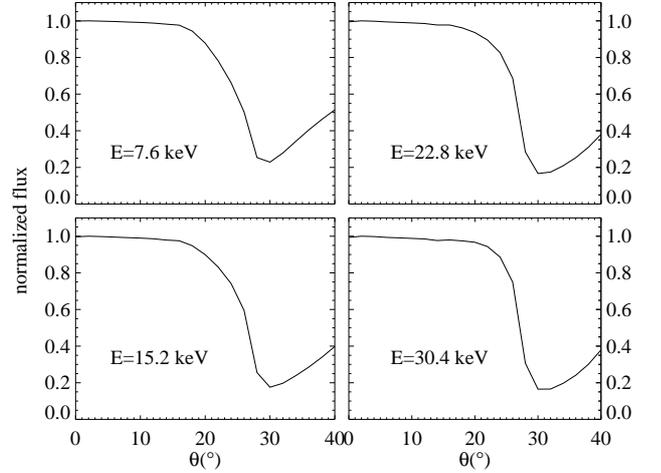}
  \caption[Computed beam pattern for different energies]
          {Computed beam patterns for different photon energies for an 
            outer opening angle $\alpha_{\mathrm{o}}=0.2$\,rad and
            a column thickness 
            $\alpha_{\mathrm{o}}-\alpha_{\mathrm{i}}=0.06$\,rad. $\theta$ has the same meaning as in Fig.~\ref{fig:bp}.}
  \label{fig:diff_en}
\end{figure}

The results of the modeled beam patterns in the $\theta\in [0\degr, 40\degr]$ 
range for different values of the inner and outer half opening angles, for a 
photon energy of $E=7.6\,$keV for a distant observer ($E=10\,$keV at the neutron star 
surface) are shown in Fig.~\ref{fig:diff_models}. 
As in Fig. ~\ref{fig:bp}, $\theta=0\degr$ means the distant observer looks down onto the magnetic pole.  
The minimum in the computed beam patterns is caused by the passage of the 
accretion stream through the line of sight. 
These modeled beam patterns reproduce the shape of the 
reconstructed beam pattern of A\,0535+26 for low values of $\theta$ well (Fig.~\ref{fig:bp}).  
In Fig.~\ref{fig:diff_models} (left) computed beam patterns for different values of 
the opening angle are shown. The position of the minimum moves to higher values of $\theta$ when the 
opening angle of the funnel increases. In Fig.~\ref{fig:diff_models} (right) computed 
beam patterns for different values of the column thickness are shown. 
The width of the minimum increases for thicker columns. 
By comparing the reconstructed beam patterns of A\,0535+26 in Fig.~\ref{fig:bp} (left) for low 
value of $\theta$ with the modeled beam patterns in Fig.~\ref{fig:diff_models}, we estimate the 
half-opening angle of the accretion stream in the case of A\,0535+26 to be $\alpha_\mathrm{o} = 0.2$~rad 
or $\sim11.5^{\circ}$ and the wall thickness to be 
$\alpha_\mathrm{o} - \alpha_\mathrm{i} = 0.06$~rad or $\sim3.4^{\circ}$. 
We used these estimates for the half-opening angle and for the wall thickness to 
investigate the energy dependence of the computed beam patterns. We calculated beam 
patterns for photon energies $E=10\,$keV, $E=20\,$keV, $E=30\,$keV, 
and $E=40$\,keV at the neutron star surface (or $E=7.6\,$keV, 
$E=15.2\,$keV, $E=22.8\,$keV, and $E=30.4\,$keV for a distant observer). 
The computed beam patterns for different photon energies are shown in 
Fig.~\ref{fig:diff_en}. The passage of the accretion stream
through the line of sight produces a distinct minimum that
only weakly depends on the photon energy as expected from a geometric signature.

\begin{table}[h]
\caption{Model parameters used in the computation of beam patterns
for A\,0535+26, for $\theta\in[0\degr, 40\degr]$.}
\label{tab:models} \centering
\begin{tabular}{|c|c|c|c|c|}\hline
model & $\alpha_{\mathrm{i}}$ (rad)& $\alpha_{\mathrm{o}}$ (rad) & $k_{\rm {B}}T_{\rm {eff}}$\,(keV)  & $\rho_{\rm {0}}$ ($10^{-5}$g\,/cm$^{-3}$)\\\hline
1    &      0.08    &  0.1  &4.1  & 16  \\
2    &      0.06    &  0.1  &4.1  & 9 \\
3    &      0.04    &  0.1  &4.1  & 6.8 \\
4    &      0.09    &  0.15 &3.7  & 4 \\
5    &      0.14    &  0.2  &3.5  & 2.8 \\\hline
\end{tabular}
\end{table}

\section{Summary and conclusions}
\label{sec:summary}
In this work, a decomposition analysis was applied to the A\,0535+26 
energy-dependent pulse profiles. A dipole magnetic field is assumed with 
axisymmetric emission regions. The asymmetry in the total pulse profiles is 
explained with a small offset from one of the magnetic poles from the 
antipodal position. We find a physically acceptable decomposition of the 
pulse profiles that allows us to extract information on the geometry of the 
pulsar. We obtain $\Theta_{1}\approx50^{\circ}$ and 
$\Theta_{2}\approx130^{\circ}$ for the position of the magnetic poles, and 
an offset of $\delta\approx25^{\circ}$. 

The visible section of the beam pattern was reconstructed. 
A characteristic feature of the reconstructed beam pattern at all energies
is a minimum observed in the flux between $\theta\approx30^{\circ}-40^{\circ}$, 
where $\theta$ is the angle between the direction of observation and the 
magnetic axis. This was interpreted in terms of a simple geometrical 
model that includes relativistic light deflection. The model 
includes a hollow column emitting isotropically black body radiation, 
plus a thermal halo created on the neutron star surface around the column 
from scattered radiation emitted from the column walls. Another 
characteristic feature of the reconstructed beam pattern is
a steep increase in flux at high values of $\theta$ ($\theta>120^{\circ}$).
This could come from gravitational light bending, which produces a 
similar feature in model calculations. 

We performed model calculations for different column thicknesses 
and opening angles, and we found the best estimates of the  half-opening angle and column 
thickness to be $\alpha_{\mathrm{o}}=0.2\,\mathrm{rad}$,  
$\alpha_{\mathrm{o}}-\alpha_{\mathrm{i}}=0.06\,\mathrm{rad}$. 
We would like to stress, however, that this model is simplified, and we do not claim 
it is true in all details, but it does reproduce the basic shape of the energy-dependent reconstructed 
beam pattern of A\,0535+26 for values of $\theta<40$\,\degr well. 
Computation of beam patterns at different energies has revealed
a weak dependence of the minimum and its depth with the energy, suggesting 
that the minimum is mainly an effect of the geometry of the system, produced
when the observer looks directly onto the accretion stream. This 
weak energy dependence on the minimum is also found in the reconstructed beam 
patterns of A\,0535+26.  

\begin{acknowledgements}
We thank the referee for providing helpful and constructive comments. 
Part of this work was supported by the Bundesministerium f\"ur Wirtschaft und Technologie 
through the German Space Agency (DLR) under contract no. 50 OR 0302. IC acknowledges both 
support and hospitality of the ESAC faculty, and financial support from the French Space Agency CNES 
through CNRS. 
\end{acknowledgements}

\bibliographystyle{aa}
\bibliography{references,a0535}

\end{document}